\documentclass[twocolumn,showpacs,preprintnumbers]{revtex4}
\usepackage{amssymb}
\usepackage{amsmath}
\usepackage{graphicx}
\usepackage{dcolumn}
\usepackage{bm}

\begin{document}

\title{Existence Criterion of Genuine Tripartite Entanglement }
\author{Chang-shui Yu}
\author{He-shan Song}
\email{hssong@dlut.edu.cn}
\affiliation{Department of Physics, Dalian University of Technology,\\
Dalian 116024, China}
\date{\today }

\begin{abstract}
In this paper, an intuitive mathematical formulation is provided to
generalize the residual entanglement for tripartite systems of
qubits (Phys. Rev. A \textbf{61}, 052306 (2000)) to the tripartite
systems in higher dimension. The spirit lies in the tensor treatment
of tripartite pure states (Phys. Rev. A \textbf{72}, 022333 (2005)).
A distinct characteristic of the present generalization is that the
formulation for higher dimensional systems is invariant under
permutation of the subsystems, hence is employed as a criterion to
test the existence of genuine tripartite entanglement. Furthermore,
the formulation for pure states can be conveniently extended to the
case of mixed states by utilizing the kronecker product approximate
technique. As applications, we give the analytic approximation of
the criterion for weakly mixed tripartite quantum states and
consider the existence of genuine tripartite entanglement of some
weakly mixed states.
\end{abstract}

 \maketitle

\section{\protect\bigskip Introduction}

Entanglement is an essential ingredient in the broad field of quantum
information theory. It is the basis of a lot of quantum protocols, such as
quantum computation [1], quantum cryptography [2], quantum teleportation
[3], quantum dense coding [4] and so on. It has been an important physical
resource. Recently, many efforts have been made on the quantification of the
resource [5,6,7,8], however, the good understanding is only limited in
low-dimensional systems. The quantification of entanglement for higher
dimensional systems and multipartite quantum systems remains to be an open
question.

Since the remarkable concurrence was presented [5], it has been
shown to be a useful entanglement measure for the systems of qubit.
More interestingly, based on the concurrence, Valerie Coffman et al
[9] introduced the so called residual entanglement for tripartite
systems of qubits. The residual entanglement is independent on the
permutation of the qubits, hence can be employed to measure genuine
three-party entanglement, i.e. the tripartite entanglement, which
opens the path to studying multipartite entanglement. Based on the
motivation of generalizing the definition of the residual
entanglement to higher dimensional systems and multipartite quantum
systems, Alexander Wong et al [10] introduced the definition of the
$n$-tangle for $n$ qubits with $n$ even, however, the $n$-tangle
itself is not a measure of the $n$-partite entanglement. Later,
hyperdeterminant in Ref. [11] has been shown to be an entanglement
monotone and represent the genuine multipartite entanglement.
However, it is easy to find that the hyperdeterminant for higher
dimensional systems and multipartite system can not be explicitly
given conveniently. In particular, so far the hyperdeterminant as an
entanglement measure has not been able to be extended to mixed
systems. Furthermore, a new method by constructing $N$-qubit
entanglement monotones was introduced by Andreas Osterloh et al [12]
for pure states to measure the $n$-partite entanglement, however, it
is only confined to the systems of qubits and seems to be very
difficult to extend to the case of mixed states analogously to Ref.
[11].

In this paper, we introduce a new approach to generalize the
residual entanglement for tripartite systems of qubits to the
tripartite systems in higher dimension. One knows that the key to
obtaining the explicit $\tau _{ABC}$ in Ref. [9] is the analytic
expression of the concurrence in mixed systems of qubits. However,
so far no one has been able to obtain an analytic expression of
concurrence (or concurrence vector) for higher dimensional mixed
systems, which means that the expectable results for higher
dimensional systems seems not to be obtained from the similar method
to that in Ref. [9]. Hence, we provide an intuitive mathematical
formulation to generalize the residual entanglement according to the
tensor treatment of tripartite pure states presented in Ref. [13]. A
distinct characteristic of the present generalization is that the
formulation for higher dimensional systems is invariant under
permutation of the subsystems (i.e. the qudits), hence can be
employed as a criterion to test existence of the genuine tripartite
entanglement (also called tripartite entanglement for convenience in
the paper). Furthermore, the formulation for pure states can be
conveniently extended to the case of mixed states by utilizing the
kronecker product approximate technique [14,15]. However, it should
be noted that the formulation is not an entanglement measure except
that for tripartite systems of qubits due to the variance under
local unitary operations. As applications, we give the analytic
approximation of the criterion for weakly mixed tripartite quantum
states (quasi pure states) and consider the existence of tripartite
entanglement of some quasi pure states, which shows that our
criterion can be conveniently applied in these cases. The paper is
organized as follows. Firstly, we give the intuitive generalization
of the residual entanglement for pure states; secondly, we extend it
to mixed states and discuss the existence of tripartite entanglement
of some quasi pure states; the conclusions are drawn in the end.

\section{Existence criterion of tripartite entanglement for pure states}

The residual entanglement for tripartite systems of qubits or $\tau _{ABC}$
(i.e. the tripartite entanglement measure) is given by

\begin{equation}
\tau (\left\vert \psi _{ABC}\right\rangle )=\sqrt{\det R}=\left\vert
d_{1}-2d_{2}+4d_{3}\right\vert ,
\end{equation}%
where a constant factor is neglected and the element $R_{ij}$ of the
$2\times 2$ matrix $R$ is defined by
\begin{equation}
R_{ij}=\sum a_{klj}a_{mni}^{\ast }\epsilon _{mp}\epsilon _{nq}a_{pqr}^{\ast
}a_{str}\epsilon _{sk}\epsilon _{tl},
\end{equation}%
with the sum being over all the repeated indices, $\epsilon _{01}=-\epsilon
_{10}=1$ and $\epsilon _{00}=-\epsilon _{11}=1$;
\begin{equation*}
d_{1}=a_{000}^{2}a_{111}^{2}+a_{001}^{2}a_{110}^{2}+a_{010}^{2}a_{101}^{2}+a_{100}^{2}a_{011}^{2};
\end{equation*}%
\begin{eqnarray*}
d_{2} &=&a_{000}a_{111}a_{011}a_{100}+a_{000}a_{111}a_{101}a_{010} \\
&&+a_{000}a_{111}a_{110}a_{001}+a_{011}a_{100}a_{101}a_{010} \\
&&+a_{011}a_{100}a_{110}a_{001}+a_{101}a_{010}a_{110}a_{001};
\end{eqnarray*}%
\begin{equation}
d_{3}=a_{000}a_{110}a_{101}a_{011}+a_{111}a_{001}a_{010}a_{100}.
\end{equation}%
What's more, the $a$ terms in above equations are the coefficients in the
standard basis defined by $\left\vert \psi _{ABC}\right\rangle
=\sum_{i,j,k=0}^{1}a_{ijk}\left\vert ijk\right\rangle _{ABC}$. As mentioned
in Ref. [9], the expression of $\tau (\left\vert \psi _{ABC}\right\rangle )$
can be mentally pictured by imagining the eight coefficients $a_{ijk}$
attached to the corners of a cube. The picture yields that $\tau $ is
invariant under permutations of the qubits, because a permutation of qubits
corresponds to a reflection or rotation of the cube. It happens that the
picture is consistent to the tensor cube introduced in Ref. [13]. In other
words, a tensor cube of $\left\vert \psi _{ABC}\right\rangle $ corresponds
to a tripartite entanglement measure $\tau (\left\vert \psi
_{ABC}\right\rangle )$. For convenience, we employ $f\left( \left\vert \psi
_{ABC}\right\rangle \right) =\left\vert \tau (\left\vert \psi
_{ABC}\right\rangle )\right\vert ^{2}$ to measure tripartite entanglement,
which is equivalent to $\tau (\left\vert \psi _{ABC}\right\rangle )$ from
the viewpoint of entanglement measure. Obviously, $f\left( \left\vert \psi
_{ABC}\right\rangle \right) $ has the same properties to $\tau (\left\vert
\psi _{ABC}\right\rangle )$.

According to Ref. [13], a tripartite pure state in any dimension can be
regarded as the tensor grid which includes tensor cubes. E.g. let $%
\left\vert \phi _{ABC}\right\rangle
=\sum_{i,j=0}^{1}\sum_{k=0}^{2}a_{ijk}\left\vert ijk\right\rangle
_{ABC}$, the tensor grid of $\left\vert \phi _{ABC}\right\rangle $
can be pictured as figure 1, which includes three tensor cubes. In
this sense, one can draw a conclusion that tensor cube is the unit
of tensor grid. Since every tensor cube in a tensor grid can be
considered as an non-normalized tripartite pure state of qubits, one
can get that every unit corresponds to the tripartite entanglement
measure of the non-normalized pure state. Namely, the tensor cube
corresponds to the minimal unit of describing the tripartite
entanglement. Therefore, whether there exist some genuine tripartite
entanglement can be determined by all the minimal units.
\begin{figure}[tbp]
\includegraphics[width=9.5cm]{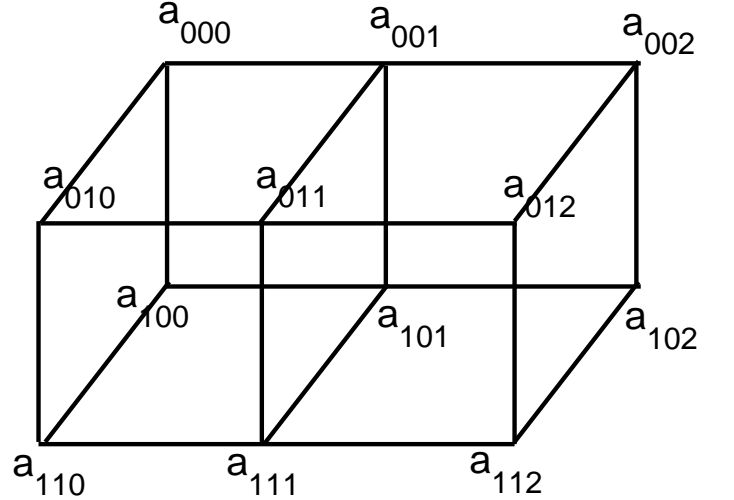}
\caption{The tensor grid of the coefficients of a tripartite pure state in $%
2\times 2\times 3$ dimension.}
\label{1}
\end{figure}

\textbf{Theorem 1:-}For any a tripartite pure state $\left\vert \Psi
\right\rangle $ which includes $M$ minimal units mentioned above,
let the the non-normalized tripartite pure state of qubits
corresponding to the $i$th unit be denoted by $\left\vert \varphi
_{i}\right\rangle $, then the corresponding tripartite entanglement
can be given by $f\left( \left\vert \varphi _{i}\right\rangle
\right) $. Define
\begin{equation}
F(\left\vert \Psi \right\rangle )=\sqrt[4]{\sum_{i=1}^{M}f\left( \left\vert
\varphi _{i}\right\rangle \right) },
\end{equation}%
for the state $\left\vert \Psi \right\rangle $, then if there does not exist
genuine tripartite entanglement in $\left\vert \Psi \right\rangle $, $%
F(\left\vert \Psi \right\rangle )=0$.

\textbf{Proof.} It is obvious that $F(\left\vert \Psi \right\rangle )=0$
means that $f\left( \left\vert \varphi _{i}\right\rangle \right) =0$ holds
for all $\varphi _{i}$, \textit{vice versa}. Since the tensor cube
corresponds to the minimal unit of describing the tripartite entanglement, $%
F(\left\vert \Psi \right\rangle )=0$ shows that there does not exist genuine
three-party entanglement in $\left\vert \Psi \right\rangle $. That is to
say, $F(\left\vert \Psi \right\rangle )$ can effectively test the existence
of tripartite entanglement in $\left\vert \Psi \right\rangle $. Furthermore,
a permutation of qudits corresponds to a reflection or rotation of the
tensor grid, which is similar to that in Ref. [9], hence all the tensor
cubes in the tensor grid are invariant except the relative positions in the
grid. Namely, $F(\left\vert \Psi \right\rangle )$ is invariant under
permutations of the qudits.

Considering the matrix notation of $\left\vert \Psi \right\rangle
=\sum_{i=0}^{n_{1}-1}\sum_{j=0}^{n_{2}-1}\sum_{k=0}^{n_{3}-1}a_{ijk}\left%
\vert ijk\right\rangle $, $F(\left\vert \Psi \right\rangle )$ can be
expressed as the function of $\left\vert \Psi \right\rangle $, i.e.%
\begin{equation}
F(\left\vert \Psi \right\rangle )=\sqrt[4]{\sum_{\alpha
=1}^{N_{1}}\sum_{\beta =1}^{N_{2}}\sum_{\gamma =1}^{N_{3}}f\left( \left(
s_{\alpha }\otimes s_{\beta }\otimes s_{\gamma }\right) \left\vert \Psi
\right\rangle \right) },
\end{equation}%
where $N_{p}=\frac{n_{p}(n_{p}-1)}{2}$ with $p=1,2,3$; $s_{q}$, $q=\alpha
,\beta ,\gamma ,$ denotes $2\times n_{p}$ matrix with $p$ corresponding to $%
q $. If the generator of the group $SO(n_{p})$ is denoted by $S_{p}$, $s_{q}$
can be derived from $\left\vert S_{p}\right\vert $ by deleting the row where
all the elements are zero, where $\left\vert \text{ }\right\vert $ denotes
the absolute value of the matrix elements.

Because eq. (2) can also be written in the standard basis by
\begin{eqnarray}
R_{ij}
&=&%
\sum_{r=0}^{1}(a_{00j}a_{11r}+a_{11j}a_{00r}-a_{01j}a_{10r}-a_{10j}a_{01r})
\notag \\
&&\cdot (a_{00i}^{\ast }a_{11r}^{\ast }+a_{11i}^{\ast }a_{00r}^{\ast
}-a_{01i}^{\ast }a_{10r}^{\ast }-a_{10i}^{\ast }a_{01r}^{\ast }),
\end{eqnarray}%
and $f(\left\vert \psi _{ABC}\right\rangle )=\det R$, $F(\left\vert \Psi
\right\rangle )$ can be expanded by
\begin{eqnarray}
&&F(\left\vert \Psi \right\rangle )=\{\sum_{\alpha =1}^{N_{1}}\sum_{\beta
=1}^{N_{2}}\sum_{\gamma =1}^{N_{3}}  \notag \\
&&[\sum\limits_{k=0}^{1}\left\vert \left\langle \Psi ^{\ast }\right\vert
S_{\alpha \beta \gamma }^{T}\left\vert \left\vert 0\right\rangle
\right\rangle (\sigma _{y}\otimes \sigma _{y})\left\langle \left\langle
k\right\vert \right\vert S_{\alpha \beta \gamma }\left\vert \Psi
\right\rangle \right\vert ^{2}  \notag \\
&&\times \sum\limits_{k=0}^{1}\left\vert \left\langle \Psi \right\vert
S_{\alpha \beta \gamma }^{T}\left\vert \left\vert 1\right\rangle
\right\rangle (\sigma _{y}\otimes \sigma _{y})\left\langle \left\langle
k\right\vert \right\vert S_{\alpha \beta \gamma }\left\vert \Psi ^{\ast
}\right\rangle \right\vert ^{2}  \notag \\
&&-\sum\limits_{k=0}^{1}(\left\langle \Psi ^{\ast }\right\vert S_{\alpha
\beta \gamma }^{T}\left\vert \left\vert 0\right\rangle \right\rangle (\sigma
_{y}\otimes \sigma _{y})\left\langle \left\langle k\right\vert \right\vert
S_{\alpha \beta \gamma }\left\vert \Psi \right\rangle  \notag \\
&&\times \left\langle \Psi \right\vert S_{\alpha \beta \gamma
}^{T}\left\vert \left\vert 1\right\rangle \right\rangle (\sigma _{y}\otimes
\sigma _{y})\left\langle \left\langle k\right\vert \right\vert S_{\alpha
\beta \gamma }\left\vert \Psi ^{\ast }\right\rangle )  \notag \\
&&\times \sum\limits_{k=0}^{1}(\left\langle \Psi ^{\ast }\right\vert
S_{\alpha \beta \gamma }^{T}\left\vert \left\vert 1\right\rangle
\right\rangle (\sigma _{y}\otimes \sigma _{y})\left\langle \left\langle
k\right\vert \right\vert S_{\alpha \beta \gamma }\left\vert \Psi
\right\rangle  \notag \\
&&\times \left\langle \Psi \right\vert S_{\alpha \beta \gamma
}^{T}\left\vert \left\vert 0\right\rangle \right\rangle (\sigma _{y}\otimes
\sigma _{y})\left\langle \left\langle k\right\vert \right\vert S_{\alpha
\beta \gamma }\left\vert \Psi ^{\ast }\right\rangle )]\}^{1/4},
\end{eqnarray}%
where $S_{\alpha \beta \gamma }=s_{\alpha }\otimes s_{\beta }\otimes
s_{\gamma }$, $\left\vert \left\vert 0\right\rangle \right\rangle =\left(
\begin{array}{c}
1 \\
0%
\end{array}%
\right) \otimes \left(
\begin{array}{cc}
1 & 0 \\
0 & 1%
\end{array}%
\right) \otimes \left(
\begin{array}{cc}
1 & 0 \\
0 & 1%
\end{array}%
\right) $, $\left\vert \left\vert 1\right\rangle \right\rangle =\left(
\begin{array}{c}
0 \\
1%
\end{array}%
\right) \otimes \left(
\begin{array}{cc}
1 & 0 \\
0 & 1%
\end{array}%
\right) \otimes \left(
\begin{array}{cc}
1 & 0 \\
0 & 1%
\end{array}%
\right) $, and $\left\langle \left\langle \ \right\vert \right\vert =\left(
\left\vert \left\vert \ \right\rangle \right\rangle \right) ^{T}$ and the
superscript $T$ denotes transposition operation. Note that $F(\left\vert
\Psi \right\rangle )=\frac{1}{2}(F(\left\vert \Psi \right\rangle
)+F(\left\vert \Psi \right\rangle )^{\ast })$. Although the expanded $%
F(\left\vert \Psi \right\rangle )$ is a bit tedious, it is important for the
extension of $F(\left\vert \Psi \right\rangle )$ to mixed states.

\section{Existence criterion of tripartite entanglement for mixed states}

\subsection{Kronecker product approximation technique}

We first introduce the kronecker product approximation technique [14,15].
For any a matrix $M=[m_{ij}]$, with entries $m_{ij}$, defined in $%
C_{d_{1}}\otimes C_{d_{2}}$, $\tilde{M}$ can be defined [16] by%
\begin{equation}
\tilde{M}=V_{12}^{L}(MV_{12}^{R})^{T_{2}},
\end{equation}%
where the superscript $T_{2}$ denotes partial transposition on the second
space [17], $V_{12}^{L,R}$ are left (right) hand side swap operators defined
as $V_{12}=\sum_{ikj^{\prime }k^{\prime }}\delta _{jk^{\prime }}\delta
_{j^{\prime }k}\left\vert j\right\rangle \left\langle j^{\prime }\right\vert
\otimes \left\vert k\right\rangle \left\langle k^{\prime }\right\vert $, $%
j,k^{\prime }=1,\cdot \cdot \cdot ,d_{2}$, $j^{\prime },k=1,\cdot \cdot
\cdot ,d_{1}$. The right hand side swap operator is defined in $%
C_{d_{1}}\otimes C_{d_{2}}$ and the left one is defined in $C_{d_{2}}\otimes
C_{d_{1}}$. Furthermore, $V_{12}^{L}=(V_{12}^{R})^{T}=(V_{12}^{R})^{-1}$. If
$d_{1}=d_{2}$, $V_{12}^{L}=V_{12}^{R}$. $\tilde{M}$ has the singular value
decompositions:%
\begin{equation}
\tilde{M}=U\Sigma V^{\dag }=\sum_{i=1}^{r}\sigma _{i}u_{i}v_{i}^{\dag },
\end{equation}%
where $u_{i}$, $v_{i}$ are the $i$th columns of the unitary matrices $U$ and
$V$, respectively; $\Sigma $ is a diagonal matrix with elements $\sigma _{i}$
decreasing for $i=1,\cdot \cdot \cdot ,r$; $r$ is the rank of $\tilde{M}$.
Based on Ref. [14,15], $M$ can be written by
\begin{equation}
M=\sum_{i=1}^{r}\left( X_{i}\otimes Y_{i}\right) ,
\end{equation}%
with $Vec(X_{i})=\sqrt{\sigma _{i}}u_{i}$ and $Vec(Y_{i})=\sqrt{\sigma _{i}}%
v_{i}^{\ast }$, where
\begin{equation}
Vec(A)=[a_{11},\cdot \cdot \cdot ,a_{p1},a_{12},\cdot \cdot \cdot
,a_{p2},\cdot \cdot \cdot ,a_{1q},\cdot \cdot \cdot ,a_{pq}]^{T},
\end{equation}%
for any a $p\times q$ matrix $A=[a_{ij}]$ with entries $a_{ij}$ [18].

\subsection{Extension of existence criterion to mixed states}

Consider $F(\left\vert \Psi \right\rangle )$ of pure states, the
corresponding quantity of mixed states $\rho $ is then given as the convex
of
\begin{equation}
F(\rho )=\inf \sum_{i}p_{i}F(\left\vert \Psi _{i}\right\rangle )
\end{equation}%
of all possible decompositions into pure states $\left\vert \Psi
_{i}\right\rangle $ with
\begin{equation}
\rho =\sum_{i}p_{i}\left\vert \Psi _{i}\right\rangle \left\langle \Psi
_{i}\right\vert ,p_{i}\geq 0.
\end{equation}%
$F(\rho )$ vanishes if and only if $\rho $ does not include any genuine
three-party entanglement. According to the matrix notation [7] of equation
(13), one can obtain $\rho =\Psi W\Psi ^{\dagger }$, where $W$ is a diagonal
matrix with $W_{ii}=p_{i}$, the columns of the matrix $\Psi $ correspond to
the vectors $\left\vert \Psi _{i}\right\rangle $. Due to the eigenvalue
decomposition: $\rho =\Phi M\Phi ^{\dagger }$, where $M$ is a diagonal
matrix whose diagonal elements are the eigenvalues of $\rho $, and $\Phi $
is a unitary matrix whose columns are the eigenvectors of $\rho $, one can
obtain $\Psi W^{1/2}=\Phi M^{1/2}U$, where $U\in C^{r\times N}$ is a
Right-unitary matrix, with $N$ and $r$ being the column number of $\Psi $
and the rank of $\rho $. Therefore, based on the matrix notation and \ eq.
(7), eq. (12) can be rewritten as%
\begin{equation*}
F(\rho )=\inf_{U}\sum_{i}^{N}([\left( U^{T}\otimes U^{\dag }\otimes
U^{T}\otimes U^{\dag }\right)
\end{equation*}%
\begin{equation}
\times \sum_{\alpha \beta \gamma }A_{\alpha \beta \gamma }\left( U\otimes
U^{\ast }\otimes U\otimes U^{\ast }\right) ]_{ii,ii}^{ii,ii})^{1/4},
\end{equation}%
where
\begin{equation*}
A_{\alpha \beta \gamma }=\frac{1}{2}\left( \mathbf{\rho }^{1/2}\right) ^{T}%
\mathbf{S}_{\alpha \beta \gamma }^{T}\left\vert \left\vert L\right\rangle
\right\rangle \Sigma _{y}\left\langle \left\langle R\right\vert \right\vert
\mathbf{S}_{\alpha \beta \gamma }\left( \mathbf{\rho }^{1/2}\right)
\end{equation*}%
defined in $C_{d\times d}\otimes C_{d\times d}\otimes C_{d\times d}\otimes
C_{d\times d}$, and $\rho $ is defined in $C_{d\times d}$, with%
\begin{equation*}
\mathbf{\rho }^{1/2}=\left( \Phi M^{1/2}\right) ^{T}\otimes \left( \Phi
M^{1/2}\right) ^{\dag }\otimes \left( \Phi M^{1/2}\right) ^{T}\otimes \left(
\Phi M^{1/2}\right) ^{\dag },
\end{equation*}

\begin{equation*}
\left\vert \left\vert L\right\rangle \right\rangle =\left\vert \left\vert
0011\right\rangle \right\rangle +\left\vert \left\vert 1100\right\rangle
\right\rangle -\left\vert \left\vert 0110\right\rangle \right\rangle
-\left\vert \left\vert 1001\right\rangle \right\rangle ,
\end{equation*}%
\begin{equation*}
\left\vert \left\vert R\right\rangle \right\rangle =(\left\langle
\left\langle 00\right\vert \right\vert +\left\langle \left\langle
11\right\vert \right\vert )\otimes (\left\langle \left\langle 00\right\vert
\right\vert +\left\langle \left\langle 11\right\vert \right\vert ),
\end{equation*}%
\begin{equation*}
\Sigma _{y}=\otimes _{j=1}^{8}\sigma _{y},
\end{equation*}%
\begin{equation*}
\mathbf{S}_{\alpha \beta \gamma }=\otimes _{j=1}^{4}S_{\alpha \beta \gamma },
\end{equation*}%
and%
\begin{equation*}
\left\vert \left\vert abcd\right\rangle \right\rangle =\left\vert \left\vert
a\right\rangle \right\rangle \otimes \left\vert \left\vert b\right\rangle
\right\rangle \otimes \left\vert \left\vert c\right\rangle \right\rangle
\otimes \left\vert \left\vert d\right\rangle \right\rangle .
\end{equation*}%
If the former two subspaces and the latter two ones are regarded as a
doubled subspace, respectively. $\sum_{\alpha \beta \gamma }A_{\alpha \beta
\gamma }$ can be considered to be defined in $C_{d^{2}\times d^{2}}\otimes
C_{d^{2}\times d^{2}}$. It is easy to find that $\sum_{\alpha \beta \gamma
}A_{\alpha \beta \gamma }$ is invariant under the exchange of two doubled
subspaces. Hence, based on the kronecker product approximation technique, $%
\sum_{\alpha \beta \gamma }A_{\alpha \beta \gamma }$ can be written by%
\begin{equation}
\sum_{\alpha \beta \gamma }A_{\alpha \beta \gamma }=\sum_{i}^{r^{\prime
}}B_{i}\otimes B_{i}=\sum_{i}^{r^{\prime }}\sigma _{i}B_{i}^{\prime }\otimes
B_{i}^{\prime },
\end{equation}%
with $B_{i}$, $B_{i}^{\prime }$ defined in $C_{d\times d}\otimes C_{d\times
d}$ and $\sigma _{i}$ the corresponding singular value. $B_{i}=\sqrt{\sigma
_{i}}B_{i}^{\prime }$ which can be obtained following the procedure in above
subsection is not given explicitly. Furthermore, $r^{\prime }$ is the rank
of the matrix $\sum_{\alpha \beta \gamma }\tilde{A}_{\alpha \beta \gamma }$
defined in above subsection. Due to eq. (15), eq. (14) can be rewritten by%
\begin{equation}
F(\rho )=\inf_{U}\sum_{i}^{N}\left( \sum_{j}^{r^{\prime }}\left( \left[
\left( U^{T}\otimes U^{\dag }\right) B_{j}\left( U\otimes U^{\ast }\right) %
\right] _{ii}^{ii}\right) ^{2}\right) ^{1/4}.
\end{equation}

It is also obvious that $A_{\alpha \beta \gamma }$ is converted into $%
A_{\alpha \beta \gamma }^{\ast }$, if the former two subspaces and the
latter two ones are exchanged simultaneously. Based on the kronecker product
approximation technique again, one can obtain that
\begin{equation*}
B_{j}=\sum_{i}^{r^{\prime \prime }}\left( C_{j}\right) _{i}\otimes \left(
C_{j}\right) _{i}^{\ast }=\sum_{i}^{r^{\prime \prime }}\left( \sigma
_{j}^{\prime }\right) _{i}\left( C_{j}^{\prime }\right) _{i}\otimes \left(
C_{j}^{\prime }\right) _{i}^{\ast }
\end{equation*}%
holds for any $j$, with $\left( C_{j}\right) _{i}$, $\left( C_{j}^{\prime
}\right) _{i}$ defined in $C_{d\times d}$, $\sigma _{j}^{\prime }$ the
corresponding singular value and $\left( C_{j}\right) _{i}=\sqrt{\left(
\sigma _{j}^{\prime }\right) _{i}}\left( C_{j}^{\prime }\right) _{i}$.
Analogously, $r^{\prime \prime }$ is the rank of $\tilde{B}_{j}$. Hence, eq.
(16) can be rewritten by%
\begin{equation}
F(\rho )=\inf_{U}\sum_{i}^{N}\left( \sum_{j}^{r^{\prime }}\left(
\sum_{m}^{r^{\prime \prime }}\left\vert \left( U^{T}\left( C_{j}\right)
_{m}U\right) _{ii}\right\vert ^{2}\right) ^{2}\right) ^{1/4}.
\end{equation}%
The infimum can be employed to test the existence of tripartite entanglement
of $\rho $.

In terms of the Cauchy-Schwarz inequality $\left(
\sum\limits_{i}x_{i}^{2}\right) ^{1/2}\left( \sum\limits_{i}y_{i}^{2}\right)
^{1/2}\geqslant \sum\limits_{i}x_{i}y_{i}$ and $\sum_{i}\left\vert
x_{i}\right\vert \geq \left\vert \sum_{i}x_{i}\right\vert $, $F(\rho )$ can
also be expressed by
\begin{equation}
F(\rho )\geq \inf_{U}\sum_{i}^{N}\left\vert U^{T}\left( \sum_{j}^{r^{\prime
}}\sum_{m}^{r^{\prime \prime }}z_{j}\cdot Z_{jm}\left( C_{j}\right)
_{m}\right) U\right\vert _{ii},
\end{equation}%
where $z_{j}=x_{j}\exp (i\phi _{j})$, with $x_{j}\geq 0$, $%
\sum_{j}x_{j}^{4}=1,$ and $Z_{jm}=y_{jm}\exp (i\varphi _{jm})$, with $%
y_{jm}\geq 0$, $\sum_{jm}y_{jm}^{2}=1$. Eq. (18) has the similar form to
that in Ref. [7], even though it is a little more complex. Therefore the
infimum of eq. (18) can be given by $\underset{\mathbf{z},\mathbf{Z}}{\text{%
max}}$ $\lambda _{1}(\mathbf{z},\mathbf{Z})-\sum_{i>1}\lambda _{i}(\mathbf{z}%
,\mathbf{Z})$, where $\lambda _{j}(\mathbf{z},\mathbf{Z})$ are the singular
values of $\left( \sum_{j}^{r^{\prime }}\sum_{m}^{r^{\prime \prime
}}z_{j}\cdot Z_{jm}\left( C_{j}\right) _{m}\right) $ in decreasing order
[7], with $\mathbf{z}=[z_{1},z_{2},\cdot \cdot \cdot ,z_{r^{\prime }}]$ and $%
\mathbf{Z=}[Z_{11},Z_{12},\cdot \cdot \cdot ,Z_{1r^{\prime \prime
}},Z_{21},\cdot \cdot \cdot ,Z_{r^{\prime }r^{\prime \prime }}]$.

In terms of the inequality $\sum_{i=1}^{n}\left\vert x_{i}\right\vert
^{2}\geq \frac{1}{n}\left( \sum_{i=1}^{n}\left\vert x_{i}\right\vert \right)
^{2}$, a more analogous form about eq. (17) to that in Ref. [7] can be given
by
\begin{equation}
F(\rho )\geq \inf_{U}\left( \frac{1}{r^{\prime }}\right)
^{1/4}\sum_{i}^{N}\left\vert U^{T}\left( \sum_{j}^{r^{\prime
}}\sum_{m}^{r^{\prime \prime }}Z_{jm}\left( C_{j}\right) _{m}\right)
U\right\vert _{ii},
\end{equation}%
The infimum can be obtained by $\left( \frac{1}{r^{\prime }}\right)
^{1/4}\times \left( \underset{\mathbf{Z}}{\text{max}}\lambda _{1}(\mathbf{Z}%
)-\sum_{i>1}\lambda _{i}(\mathbf{Z})\right) $, where $\lambda _{j}(\mathbf{Z}%
)$ are the singular values of $\left( \sum_{j}^{r^{\prime
}}\sum_{m}^{r^{\prime \prime }}Z_{jm}\left( C_{j}\right) _{m}\right) $ in
decreasing order. Both the two cases can provide the necessary condition for
the existence of tripartite entanglement of a mixed state, but the
sufficiency of them may be different.

What's more, compared with the procedure in Ref. [8], it is very possible
that $\left[ \left( C_{j}\right) _{m}\right] _{\max }$ corresponding to the
maximal $\sqrt{(\sigma _{j}^{\prime })_{m}}$ can give the main contribution
to the infimum of eq. (18). That is to say the lower bound of $F(\rho )$ can
be given by $\lambda _{1}-\sum_{i>1}\lambda _{i}$ with $\lambda _{j}$ the
singular values of $\left[ \left( C_{j}\right) _{m}\right] _{\max }$.

\subsection{Examples}

In above subsection, we have provided three different lower bounds
for any mixed state, which can be employed as necessary conditions
to test the existence of tripartite entanglement in principal.
However, by analysis, one can find that the numerical realization to
calculate the bounds for a mixed state $\rho $ requires the
eigenvalue decomposition of a matrix defined in the same dimension
to that of $\otimes _{i=1}^{4}\rho $ , which reduces the efficiency
of calculation. In order to avoid the similar problem, an analytic
approximation method was introduced for quasi pure states in Ref.
[19]. By utilizing the analogous method, one will find that eq. (17)
can be simplified significantly, hence our criterion can work well
for quasi pure states. Before the examples, we firstly give the
analytic approximation of eq. (17).

Let $\sum_{\alpha \beta \gamma }A_{\alpha \beta \gamma }$ in eq.
(14) be denoted by $A$. Analogous to Ref. [19], the tensor $A$ can
be obtained by

\begin{eqnarray}
&&A_{l^{\prime }m^{\prime },j^{\prime }k^{\prime }}^{lm,jk}  \notag \\
&=&\sum_{\alpha =1}^{N_{1}}\sum_{\beta =1}^{N_{2}}\sum_{\gamma =1}^{N_{3}}[%
\sqrt{u_{l}u_{m}u_{j}u_{k}u_{l^{\prime }}u_{m^{\prime }}u_{j^{\prime
}}u_{k^{\prime }}}  \notag \\
&&\times \sum_{i=0}^{1}(\left\langle \Psi _{l}^{\ast }\right\vert S_{\alpha
\beta \gamma }^{T}\left\vert \left\vert 0\right\rangle \right\rangle (\sigma
_{y}\otimes \sigma _{y})\left\langle \left\langle i\right\vert \right\vert
S_{\alpha \beta \gamma }\left\vert \Psi _{l^{\prime }}\right\rangle  \notag
\\
&&\times \left\langle \Psi _{m}\right\vert S_{\alpha \beta \gamma
}^{T}\left\vert \left\vert 0\right\rangle \right\rangle (\sigma _{y}\otimes
\sigma _{y})\left\langle \left\langle i\right\vert \right\vert S_{\alpha
\beta \gamma }\left\vert \Psi _{m^{\prime }}^{\ast }\right\rangle )  \notag
\\
&&\times \sum_{i=0}^{1}(\left\langle \Psi _{j}^{\ast }\right\vert S_{\alpha
\beta \gamma }^{T}\left\vert \left\vert 1\right\rangle \right\rangle (\sigma
_{y}\otimes \sigma _{y})\left\langle \left\langle i\right\vert \right\vert
S_{\alpha \beta \gamma }\left\vert \Psi _{j^{\prime }}\right\rangle  \notag
\\
&&\times \left\langle \Psi _{k}\right\vert S_{\alpha \beta \gamma
}^{T}\left\vert \left\vert 1\right\rangle \right\rangle (\sigma _{y}\otimes
\sigma _{y})\left\langle \left\langle i\right\vert \right\vert S_{\alpha
\beta \gamma }\left\vert \Psi _{k^{\prime }}^{\ast }\right\rangle )  \notag
\\
&&-\sum_{i=0}^{1}(\left\langle \Psi _{l}^{\ast }\right\vert S_{\alpha \beta
\gamma }^{T}\left\vert \left\vert 1\right\rangle \right\rangle (\sigma
_{y}\otimes \sigma _{y})\left\langle \left\langle i\right\vert \right\vert
S_{\alpha \beta \gamma }\left\vert \Psi _{l^{\prime }}\right\rangle  \notag
\\
&&\times \left\langle \Psi _{m}\right\vert S_{\alpha \beta \gamma
}^{T}\left\vert \left\vert 0\right\rangle \right\rangle (\sigma _{y}\otimes
\sigma _{y})\left\langle \left\langle i\right\vert \right\vert S_{\alpha
\beta \gamma }\left\vert \Psi _{m^{\prime }}^{\ast }\right\rangle )  \notag
\\
&&\times \sum_{i=0}^{1}(\left\langle \Psi _{j}^{\ast }\right\vert S_{\alpha
\beta \gamma }^{T}\left\vert \left\vert 0\right\rangle \right\rangle (\sigma
_{y}\otimes \sigma _{y})\left\langle \left\langle i\right\vert \right\vert
S_{\alpha \beta \gamma }\left\vert \Psi _{j^{\prime }}\right\rangle  \notag
\\
&&\times \left\langle \Psi _{k}\right\vert S_{\alpha \beta \gamma
}^{T}\left\vert \left\vert 1\right\rangle \right\rangle (\sigma
_{y}\otimes \sigma _{y})\left\langle \left\langle i\right\vert
\right\vert S_{\alpha \beta \gamma }\left\vert \Psi _{k^{\prime
}}^{\ast }\right\rangle )],
\end{eqnarray}%
where $\Psi _{\alpha }$ denotes the $\alpha $th eigenvector and all the
other quantities are defined similar to those in eq. (7). According to the
symmetry of $A$ and the kronecker product approximation technique in above
section, $A$ can be formally written as
\begin{equation*}
A_{l^{\prime }m^{\prime },j^{\prime }k^{\prime }}^{lm,jk}=\sum_{\alpha
}T_{lm}^{\alpha }\left( T_{l^{\prime }m^{\prime }}^{\alpha }\right) ^{\ast
}T_{jk}^{\alpha }\left( T_{j^{\prime }k^{\prime }}^{\alpha }\right) ^{\ast }.
\end{equation*}%
The density matrix of quasi pure states has one single eigenvalue $\mu _{1}$
that is much larger than all the others, which induces a natural order in
terms of the small eigenvalues $\mu _{i}$, $i>1$. Due to the same reasons to
those in Ref. [19], here we consider the second order elements of type $%
A_{11,11}^{lm,11}$. Therefore, one can have the approximation
\begin{equation*}
A_{l^{\prime }m^{\prime },j^{\prime }k^{\prime }}^{lm,jk}\simeq \tau
_{lm}\tau _{l^{\prime }m^{\prime }}^{\ast }\tau _{jk}\tau _{j^{\prime
}k^{\prime }}^{\ast }\text{ with }\tau _{lm}=\frac{A_{11,11}^{lm,11}}{\sqrt[4%
]{\left( A_{11,11}^{11,11}\right) ^{3}}}.
\end{equation*}%
In this sense, eq. (17) and eq. (18) can be simplified significantly:%
\begin{equation*}
F(\rho )\simeq F_{a}(\rho )=\inf_{U}\sum_{i}\left\vert U^{T}\tau
U\right\vert _{ii}.
\end{equation*}%
$F_{a}(\rho )$ can be given by
\begin{equation*}
F_{a}(\rho )=\max \{\lambda _{1}-\sum_{i>1}\lambda _{i},0\},
\end{equation*}%
where $\lambda _{i}$ is the singular value of $\tau $ in decreasing order.

The tripartite mixed states introduced in Ref .[20]
\begin{equation*}
\rho (x)=x\left\vert GHZ\right\rangle \left\langle GHZ\right\vert
+(1-x)/2(\left\vert W\right\rangle \left\langle W\right\vert +\left\vert
\tilde{W}\right\rangle \left\langle \tilde{W}\right\vert ),
\end{equation*}%
where
\begin{equation*}
\left\vert GHZ\right\rangle =\frac{1}{\sqrt{2}}(\left\vert 000\right\rangle
+\left\vert 111\right\rangle ),
\end{equation*}%
\begin{equation*}
\left\vert W\right\rangle =\frac{1}{\sqrt{3}}\left( \left\vert
001\right\rangle +\left\vert 010\right\rangle +\left\vert 100\right\rangle
\right) ,
\end{equation*}%
\begin{equation*}
\left\vert \tilde{W}\right\rangle =\frac{1}{\sqrt{3}}\left( \left\vert
110\right\rangle +\left\vert 011\right\rangle +\left\vert 101\right\rangle
\right) ,
\end{equation*}%
can be considered as a quasi pure state for $x>1/3$. $F_{a}(\rho
(x))$ is shown in Fig. 2, which indicates the consistent conclusion
to that in Ref. [20]. What's more, for the quasi pure states
generated by the mixture of maximally mixed state(identity matrix)
and tripartite GHZ state (The cases in $3\times3\times3$ dimension
is included.), the corresponding $F_{a}(\rho )$s can all be shown to
be \emph{nonzero}. Since $F_{a}(\rho )$ is not monotone in higher
dimension, the corresponding figures are not given here.

\begin{figure}[tbp]
\includegraphics[width=8.5cm]{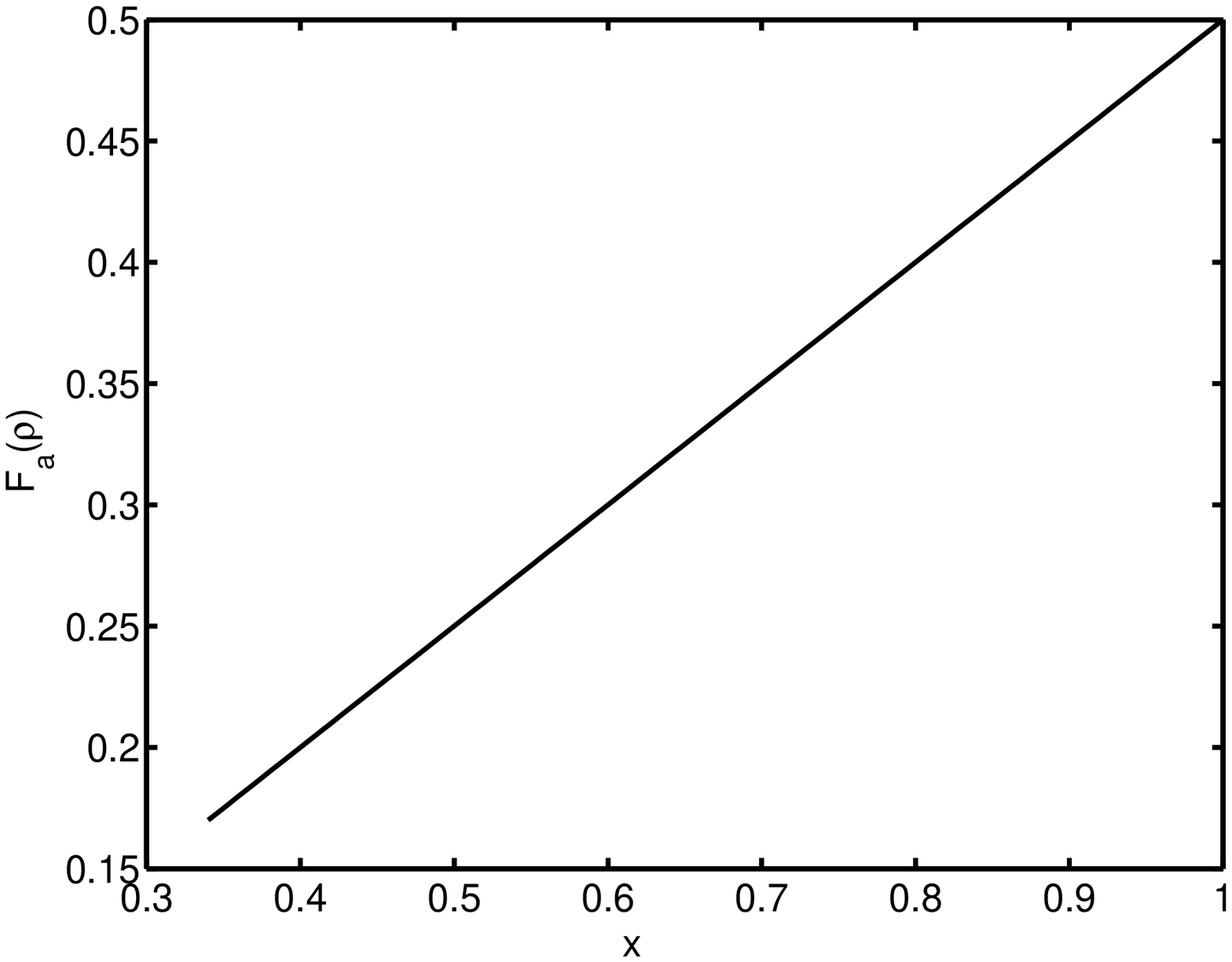}
\caption{$F_{a}(\protect\rho )$ of the mixed state $\protect\rho %
(x)=x\left\vert GHZ\right\rangle \left\langle GHZ\right\vert
+(1-x)/2(\left\vert W\right\rangle \left\langle W\right\vert +\left\vert
\tilde{W}\right\rangle \left\langle \tilde{W}\right\vert )$ \ vs $x$, $x\in
(1/3,1].$}
\end{figure}

\section{Conclusion and Discussion}

In summary, we have introduced an intuitive mathematical formulation to
generalize the original tripartite entanglement to higher dimensional
tripartite systems according to the tensor treatment of a tripartite pure
state. A distinct characteristic of the present generalization is that the
formulation for higher dimensional systems is invariant under permutation of
the qudits. When the formulation is reduced to tripartite systems of qubits,
there exists an exponent $\frac{1}{2}$ different from the original one, but
the change of exponent provides convenience for the generalization to mixed
states. The formulation for pure states can be conveniently extended to the
case of mixed states by utilizing the kronecker product approximate
technique. We have presented three different lower bounds for $F(\rho )$ of
mixed states. The forms of the three results for mixed states are similar to
those of bipartite entanglement [7,8]. All of them can provide necessary
conditions to test the existence of tripartite entanglement, but the
sufficiency of them may be different. However, because the dimension of $%
A_{\alpha \beta \gamma }$ is much higher than that corresponding to
bipartite entanglement, it seems to be a bit difficult to directly
apply to test the existence of tripartite entanglement of a general
quantum mixed state. Fortunately, for the weakly mixed states, i.e.
quasi pure states, one can find that our criterion can be
conveniently applied and is even a sufficient condition for the
existence of tripartite entanglement. In particular, our criterion
can provide an analytic approximation. Since the 3-tangle is an
entanglement measure, $F_{a}(\rho )$ is not only an existence
criterion, but also an effective tripartite entanglement indicator.
Even though there exist some questions left open, the intuitive
mathematical formulation of tripartite entanglement and the
convenient extension to mixed states will play an important role in
the further understanding of multipartite entanglement measure.

\section{Acknowledgement}

This work was supported by the National Natural Science Foundation of China,
under Grant No. 60472017.

\end{document}